\documentstyle[aps,multicol,psfig]{revtex} 

\begin{document}
\draft

\title{Spin magnetization of small metallic grains} 

\author{M. Schechter}

\address{Department of Physics and Astronomy, University of British Columbia, 
Vancouver, British Columbia, Canada V6T 1Z1}

\maketitle

%\vspace{-0.5cm}

\begin{abstract}

Small metallic grains which satisfy the conditions of the universal
Hamiltonian are considered. It is shown that for such grains the
effects of the interactions in the spin channel and in the Cooper
channel on their spin magnetization are well separated, thus allowing
the determination of the interaction parameters within this model. In
particular, the existence of pairing correlations in small grains and
the sign of the interaction in the Cooper channel can be uniquely
determined.

\end{abstract}

%\vspace{0.5cm}

\begin{multicols}{2}

\section{Introduction}

In general, the problem of disorder and interaction in electron
systems is a very difficult one. However, it was
shown\cite{AK98,BOH99,BUG00,KAA00} that for small diffusive metallic
grains with large dimensional conductance $g=E_{\rm Th}/d$ the problem
simplifies considerably. Here $d$ is the mean level spacing and $E_{\rm
Th}=\hbar D/L^2$ is the Thouless energy which is the inverse time to
diffuse across the grain. $D$ is the diffusion constant and $L$ is the
grain's size.  The low energy physics of such small grains is
described to leading order in $1/g$ by the ``universal
Hamiltonian''\cite{KAA00}, in which only the diagonal matrix elements
of the interaction survive:

\begin{equation}
H = \sum_{n=1}^{\Omega/d} \sum_{\sigma} \epsilon_n c^{\dagger}_{n,\sigma} 
c_{n,\sigma} + E_c \hat{N}^2 + J_c \hat{T}^{\dagger} \hat{T} + 
J_s \hat{S}^2. 
\label{universal}
\end{equation}

The index $n$ spans a shell of $\Omega/d$ doubly degenerate time reversed 
states of energy $\epsilon_n$, 
$\hat{N} = \sum_{n=1}^{\Omega/d} 
\sum_{\sigma}  c^{\dagger}_{n,\sigma} c_{n,\sigma}$ 
is the number operator, $\hat{\vec{S}} =\frac{1}{2}
\sum_{n=1}^{\Omega/d} \sum_{\sigma, \sigma '} c^{\dagger}_{n,\sigma}
\vec{\sigma}_{\sigma,\sigma '} c_{n,\sigma '}$ is the total spin
operator, and $\hat{T} = \sum_{n=1}^{\Omega/d} c_{n,-} c_{n,+}$ is the
pair annihilation operator. $E_c$ is the charging energy and $J_{c(s)}
= \lambda_{c(s)} d$, where $\lambda_c$ and $\lambda_s$ are the
dimensionless interaction parameters in the Cooper channel and in the
spin channel respectively. $\Omega$ is of the order of $E_{\rm Th}$,
and we take $\Omega/d = 2g$.  Recently, a similar problem of a
ballistic grain with chaotic boundary conditions was addressed using
renormalization group approach, and it was shown\cite{MM02,MS03} that
for weak interactions the low energy physics is indeed controlled by
the universal Hamiltonian.

This relatively simple description of the low energy physics of
diffusive metallic grains provides the opportunity to consider
theoretically, and eventually experimentally, problems which in bulk
systems are much harder to attack. One interesting problem is the
question of whether metals such as Gold, Copper, and Silver are
superconducting or not at very low temperatures\cite{MM00}, i.e. if
their effective interaction in the Cooper channel is attractive or
repulsive. While all these metals are not found to be superconducting
down to currently accessible temperatures, it may well be that their
effective electron-electron interaction is attractive but small. Since
$T_{\rm c}$ depends exponentially on the interaction, such weak
interaction will lead to unmeasurable $T_{\rm c}$. However, small
effective attractive interaction in such metals would affect other
properties, like the proximity effect\cite{VMP90,MM00} and persistent
currents\cite{AE90b,SOIL03}, which depend linearly on the interaction.
Furthermore, the magnitude of the effective attractive interaction in
these metals may be size dependent, as can be inferred from the
apparent size dependence of $T_c$ in many superconducting
materials\cite{ACC66,STKC70,DP74,MO84}.  In particular, Platinum,
which is not known to be a superconductor in bulk form, was recently
reported to be superconducting at very low temperatures in granular
form\cite{KSH99}.

While the determination of the effective interaction in bulk materials
is a difficult task, it was already recognized that weak pairing
correlations can be detected in small ``superconducting''
grains\cite{LFH+00,SILD01,FFM02}.  In these works it was shown that
the existence of weak pairing correlations will result in measurable
effects in the spin susceptibility\cite{LFH+00,SILD01} and specific
heat\cite{FFM02} of the grains. All these works considered the reduced
BCS Hamiltonian, in which only the pairing interaction
exists. However, in a real system other interactions exist, and in
order to experimentally determine the existence of pairing
correlations one has to show that the measured effect is uniquely
caused by the pairing interaction itself.

Small disordered metallic grains with $g \gg 1$ and not too strong
interactions\cite{MM02,MS03} are favorable from this point of view, as
they satisfy the validity conditions of the universal Hamiltonian
model, and therefore the constraints this model dictates on the
interaction terms. In this paper we calculate the ensemble averaged
differential spin susceptibility $\chi_s$ at $T=0$ of such isolated
grains, and show that the effects of the different interaction terms
are well separated, thus allowing an unequivocal determination of the
existence of pairing correlations in such grains, and furthermore, a
determination of the sign and magnitude of the effective interaction
constants as they appear in the universal Hamiltonian. Actually, we
consider the determination of $\lambda_s$ and $\lambda_c$ only. Since
the grains are isolated, the charging energy $E_{\rm c}$ is not
relevant, and could be determined by complementary tunneling
experiments. We consider the regime of $|\lambda_c|,|\lambda_s| \ll
1$. Note, that for $\lambda_c < 0$ two regimes exist, the perturbative
regime and the superconducting regime, for which $|\lambda_c| >
1/\ln{[E_{\rm Th}/d]}$\cite{SILD01}.  We first consider the former,
and then the latter regime.

\section{The perturbative regime}

Using the universal Hamiltonian, we assume that the spin-orbit
interaction is small and neglect it \cite{KAA00}. This assumption
should be verified when comparing our results with experiments,
keeping in mind the specifics of spin-orbit interaction in small
grains (see e.g. \cite{HSO+01,FA02}).  Throughout the paper we will be
interested in the ensemble averaged differential spin susceptibility
at magnetic fields $H \gg d/\mu_{\rm B}$. In this regime we can
neglect level statistics and assume that the energy levels in the
grains are equally spaced. Differences between grains with odd number
and even number of electrons can be neglected in this regime as well,
and for simplicity we consider grains with even number of
electrons. For detailed considerations regarding the neglect of level
statistics and even-odd effects see section III of
Ref.\cite{SILD01}. In particular, ensembles of the order of $10^6$
grains or larger are required for the shift in the magnetization (see
below) to be larger than the fluctuations due to level statistics.  We
also neglect orbital magnetization. This can be achieved in pancake
shaped grains (see e.g. Ref.~\cite{NGD+02}), when the field is applied
in the direction of the thin part. Practically, orbital magnetization
can not be completely avoided, but its relative magnitude can be
experimentally determined by changing the direction of the applied
magnetic field.

The spin magnetization of a grain is given by 

\begin{equation}
M = \mu_{\rm B} (n_+ - n_-) 
\label{magnetization}
\end{equation} 
where $n_+$ and $n_-$ are the number of electrons with spin parallel
and anti-parallel to the magnetic field, respectively. We define $l$ as
the number of flipped spins, such that $n_+-n_-=2l$. It can be shown
that among all states with $l$ flipped spins, the one that has the
lowest energy has all $l$ states above $E_{\rm F}$ and $l$
states below $E_{\rm F}$ singly occupied by electrons with spin
parallel to the magnetic field. The number $l$ that is realized at a
given magnetic field is the one minimizing the total energy of the
grain:

\begin{equation}
E(l) = E_0 + E^l_{\rm kin} + E^l_{\rm int} - 2 l \mu_{\rm B} H 
\, \, .
\label{El}
\end{equation}
Here $E_0$ is the energy of the noninteracting Fermi state 
(with $l=0$, no singly occupied single 
particle states), $E^l_{\rm kin} = l^2 d$ is the 
kinetic energy cost of flipping $l$ pairs, $E^l_{\rm int}$ is the energy 
due to the interaction, and $- 2 l \mu_{\rm B} H$ is the Zeeman energy. 
In order to calculate $E^l_{\rm int}$ 
we use Richardson's exact solution\cite{Ric63,RS64}. Although this solution 
was derived for the reduced BCS Hamiltonian, it can be easily generalized 
to solve the universal Hamiltonian for isolated grains. The $\hat{N}^2$ 
term is then not relevant, and the only relevant extra term in the universal 
Hamiltonian compared to the reduced BCS Hamiltonian is the spin term.

Given $l$ flipped spins, levels ${g-l+1...g+l} \equiv B$ are singly
occupied, and do not participate in the pairing
interaction\cite{detectingnote}.  Denoting $U = \Omega \setminus B$,
and neglecting the spin term, Richardson's solution is given by a set
of $k$ coupled nonlinear equations, the $\nu$'th equation of which is
given by\cite{RS64}:

\begin{equation}
-\frac{1}{\lambda_c d} +  \sum_{\mu = 1 (\neq \nu)}^k 
\frac{2}{E_\mu - E_\nu} - \sum_j^U 
\frac{1}{2 \epsilon_j - E_\nu} = 0 .
\label{Richeq}
\end{equation} 
Here $k$ is half the number of the ``paired'' electrons, and in our case 
$k=g-l$. The total energy of the system is given by  
\begin{equation}
E_{\rm BCS} = \sum_j^B \epsilon_j + \sum_{\nu=1}^k E_\nu \, , 
\label{energyeqapp}
\end{equation}
and the many-body wave function is also given in terms of the $k$
energy parameters $\{E_\nu\}$ which solve the equations (\ref{Richeq}). 
Since the electrons participating in the pairing
interaction have zero total spin, including the spin term and the Zeeman
term does not change Richardson's equations, energy parameters, and
orbital wavefunction. The spin and Zeeman terms do change the energy of
the system, for a given $l$ by $E_s=\lambda_s d l (l+1)$ and 
$E_Z= - 2 l \mu_{\rm B} H$ respectively.

The total energy can therefore be written as: 

\begin{equation}
E(l) = \sum_j^B \epsilon_j + \sum_{\nu=1}^k E_\nu + 
\lambda_s d l (l+1) - 2 l \mu_{\rm B} H\, , 
\label{energyUH}
\end{equation} 
or, in accordance with Eq.~(\ref{El})

\begin{equation}
E(l) = E_0 + l^2 d + \sum_{\nu=1}^k \delta E_\nu  
+ \lambda_s d l (l+1) - 2 l \mu_{\rm B} H\, , 
\label{energyUHeq3}
\end{equation} 
where $\delta E_\nu \equiv  E_\nu - 2 \epsilon_\nu$.
Therefore, $E_{\rm int} = \lambda_s d l (l+1) + E_{\rm pair}$ where 
\begin{equation}
E_{\rm pair} \equiv \sum_{\nu=1}^k \delta E_\nu 
\label{epair}
\end{equation}
is the energy due to the 
interaction in the Cooper channel, and the problem reduces to
finding $E_{\rm pair}(l)$. In Ref.~\cite{SILD01} this
was done to second order in the interaction
$\lambda_c$. Here we use Richardson's exact solution for the
determination of $E_{\rm pair}(l)$. This formalism allows a 
rigorous inclusion of the spin term. It also allows the possibility to
give a general expression for $E_{\rm pair}(l)$, and then obtain the
result to second order in $\lambda_c$ as an expansion of the exact
result.

Manipulating Eq.~(\ref{Richeq}) one obtains\cite{SILD01}

\begin{equation}
\delta E_\nu = \frac{\lambda_c d}{1 + \lambda_c \, a_\nu} \; , 
\label{delEnu}
\end{equation} 
where 
\begin{equation}
a_\nu = d \left( \sum_{j\neq \nu}^{U} \frac{1}{2 \epsilon_j - E_\nu} - 
\sum_{\mu=1(\neq \nu)}^k \frac{2}{E_\mu - E_\nu} \right). 
\label{anu}
\end{equation} 
For the lowest energy solution, we approximate $\delta E_\nu$ by
\begin{equation}
\delta E_\nu^0 \equiv \lambda_\nu d \; , \qquad
\mbox{where} \quad \lambda_\nu \equiv
\frac{\lambda_c }{1 + \lambda_c \, a_\nu^0} \; ,
\label{delEnuapprox}
\end{equation} 
and  $a_\nu^0 \equiv a_\nu (\lambda_c=0)$ is given by
\begin{equation}
a_\nu^0 =  \sum_{j\neq \nu}^U \frac{1}{2 j - 2 \nu} - 
\sum_{\mu=1(\neq \nu)}^k \frac{1}{\mu - \nu}.
\label{anu0}
\end{equation}
This approximation is exact to second order in $\lambda_c$, and its
accuracy to higher orders in $\lambda_c$ was studied in
Ref.~\cite{SILD01}.  $E_{\rm pair}$ can now be calculated to any order
in $\lambda_c$ by inserting expression (\ref{anu0}) in
Eq.~(\ref{epair}). To second order in $\lambda_c$ this gives

\begin{equation}
E_{\rm pair}(l) = \lambda_c d (g - l) + 
\frac{1}{2} \lambda_c^2 d \sum_{\nu=1}^{g-l} 
\ln\left[\frac{g+l+\nu}{2l + \nu}\right] \, . 
\label{epairlog}
\end{equation}

Inserting Eq.~(\ref{epairlog}) into Eq.~(\ref{energyUHeq3}) and
differentiating with respect to $l$ we obtain an equation for $l$ that
minimizes $E(l)$

\begin{equation}
2 l d + \lambda_s d (2 l + 1) - \lambda_c d + \lambda_c^2 d 
\ln\left[\frac{g}{2l}\right] - 2 \mu_{\rm B} H = 0 
\, \, , 
\label{minl}
\end{equation} 
which results in  

\begin{equation}
M=\frac{\mu_{\rm B} [2 \mu_{\rm B} H/d - 
\lambda_c^2 \ln[E_{\rm Th}/(2 \mu_{\rm B} H)] + 
\lambda_c - \lambda_s]}{1+\lambda_s} \, \, .
\label{mag}
\end{equation}
In Eqs.~(\ref{minl}) and (\ref{mag}), for the values inside the
logarithm, we assume $l \ll g$ and replace $l$ with its noninteracting
value.  The $l$ that minimizes $E(l)$ as obtained from Eq.~(\ref{minl}) is
given by the condition that the energy gain from the Zeeman term when
flipping another electron and creating 2 additional singly occupied
states with spin up electrons is equal to the energy cost of flipping
this electron, resulting from the kinetic energy, spin interaction and 
pairing interaction. The kinetic part alone produces the
noninteracting result [$\chi_0$ in Eq.~(\ref{chi}) below for the
susceptibility]. The leading contribution of the spin part to the
total energy is proportional to $l^2$, like the kinetic energy, and
this results in an effective renormalization of the density of
states. The second part of the spin term, as well as the leading part
of the pairing interaction, contribute to the total energy terms which
are linear in $l$, like the Zeeman term, and therefore result in a
constant shift of the magnetization, and do not affect $\chi_s$. 
The field dependent correction to $\chi_s$
comes from the higher orders of the pairing term, of which the second
order gives the dominant contribution. This part gives a negative
correction to the energy which is monotonically decreasing with
increasing $l$, therefore contributing a positive, field dependent
contribution to $\chi_s$. 

Differentiating with respect to $H$ we obtain the ensemble averaged
spin susceptibility for $d/\mu_{\rm B} \ll H \ll E_{\rm Th}/\mu_{\rm B}$

\begin{equation}
\chi_s=\frac{\chi_0}{1+\lambda_s} 
\left(1+ \frac{\lambda_c^2 d}{2 \mu_{\rm B} H} \right)  \, \, .
\label{chi}
\end{equation}
This is our central result. The interaction in the spin channel results 
in an $H$ independent shift of the susceptibility by a factor of 
$1/(1+\lambda_s)$. This gives the possibility to determine $\lambda_s$, by 
e.g. the Sommerfeld-Wilson ratio, that compares $\chi_s$ to the linear 
specific heat coefficient. 
The interaction in the Cooper channel results in a $1/H$
correction to $\chi_s$. This correction is a finite size effect, 
as it is proportional to the level spacing. Moreover, {\it this correction 
unequivocally signals the presence of pairing correlations in small 
metallic grains}, as it does not result from the interaction in the spin 
channel or the charging energy, and all other interactions have $1/g$ 
smallness. Interestingly, the $1/H$ correction does not depend on the sign 
of the interaction, and 
therefore exists for attractive as well as repulsive interaction in the 
Cooper channel. Thus, measuring $\chi_s$ in small metallic 
grains at magnetic fields $H \gg d/\mu_{\rm B}$ determines the magnitude of 
$\lambda_c$, but not its sign. In order to obtain the sign of $\lambda_c$ 
one has to look at $M/H$. Unlike the case in the 
susceptibility, where the first order term in the interaction is not 
field dependent, and therefore does not contribute, here, to leading 
order in $\lambda_c$ 

\begin{equation}
\frac{M}{H}=\frac{\chi_0}{1+\lambda_s} 
\left[1+ \frac{(\lambda_c-\lambda_s)d}{2 \mu_{\rm B}H}\right]  
\, \, ,
\label{MH}
\end{equation}
and the $1/H$ correction does depend on the sign of $\lambda_c$. Once
$\lambda_s$ is either known or small, the sign of $\lambda_c$ is
easily determined. Note, that in principal the information given by
$\chi_s$ and by $M/H$ is equivalent. However, their high magnetic
field behavior is different, and therefore both the sign and magnitude
of $\lambda_c$ can be obtained. (Actually, both can be obtained from
the behavior of $M/H$. However, the susceptibility measurement is
preferable for the determination of the magnitude of $\lambda_c$
because it is independent of any other interaction. It is also a more
precise measurement experimentally). The magnetic field range for
which our treatment is valid is given above Eq.(\ref{chi}), and
depends on the specific metallic grain, as well as its size and its
dimensionless conductance.  For example, for Copper grains of size
$5*50*50 {\rm nm}^3$ and $g=25$ the level spacing is roughly $0.06
{\rm K}$, the Thouless energy $1.5 {\rm K}$, and therefore the magnetic
field range would be between $0.1$ and $2.5$ Tesla.

\section{The superconducting regime}

So far we considered the perturbative regime, which for attractive 
interaction corresponds to 
$|\lambda_c| < 1/\ln{[E_{\rm Th}/d]}$ which is equivalent to $d>\Delta$ where 
$\Delta$ is the bulk gap in the mean field BCS approximation. 
In the crossover regime, 
where $d \approx \Delta$, the behavior of $\chi_s$ changes 
considerably in the low magnetic field regime, $\mu_{\rm B} H \lesssim d$. 
However, the properties of $\chi_s$ at high magnetic field 
$\mu_{\rm B} H \gg \Delta^2/d$ are similar to those in the perturbative 
regime\cite{SILD01}, and 
the interaction parameters can be similarly determined. 
The parameters of the universal Hamiltonian can also be determined in the 
``BCS regime'', where $|\lambda_c| > 1/\ln{[E_{\rm Th}/d]}$ and the level 
spacing $d \ll \Delta$ and can therefore 
be neglected. In this regime $\lambda_c$ is easy to determine, e.g. by 
measuring the excitation gap. In order to determine $\lambda_s$ in this 
regime we revisit the spin magnetization of the system. For $\lambda_s=0$ 
it is well known\cite{Clo62,Cha62} 
that the spin magnetization of a superconductor is zero 
below a value of $H=\Delta/(\sqrt{2} \mu_{\rm B})$, where a sharp 
step to the value of the spin magnetization of noninteracting electrons at 
the same $H$ occurs. The area between the magnetization curves of the 
noninteracting and superconducting systems gives the condensation energy, 
$\Delta^2/(2d)$. 
We have already shown that finite $\lambda_s$ changes the slope of the 
spin magnetization of noninteracting electrons [see Eq.~(\ref{chi}) with 
$\lambda_c=0$]. Here we show that it also changes the value of $H$ at which 
the step in the magnetization of a superconducting system occurs, as to 
keep the area between the magnetization curves to equal $\Delta^2/(2d)$. 
Thus, one can determine $\lambda_s$ in the superconducting regime by the 
magnetic field value of the magnetization step. 
This value of $H$ is where the normal and superconducting states have 
the same energy, i.e. when the equation 

\begin{equation}
l^2 d + J_s l(l+1) + \frac{\Delta^2}{2d} - 2 l \mu_{\rm B} H = 0
\end{equation}
has one solution. This occurs when $l=\Delta/\sqrt{2 d (d+J_s)}$, or when 

\begin{equation}
H = \frac{\Delta}{\sqrt{2} \mu_{\rm B}} \sqrt{1 + \lambda_s} \, \, .
\end{equation} 
The shift in the magnetic field value of the spin magnetization step
is a direct measure of $\lambda_s$ in this regime.  

\section{Summary}

We have thus shown that the determination of the interaction
parameters in small metallic grains with not too large interactions
can be done by measuring their ensemble averaged differential spin
susceptibility. Such a measurement, done systematically as function of 
grain size, can shed light on the change of transition temperature with 
grain size in granular superconductors. 
Although our theory is valid for finite size grains, and can not
directly determine if a certain material is superconducting at low
temperatures in bulk form, a systematic measurement of the interaction
parameters as function of grain size can suggest the bulk behavior as
well. 

\acknowledgments

It is a pleasure to thank Oded Agam, Joshua Folk, Yoseph Imry, Yuval
Oreg, Dror Orgad, Avraham Schiller and Alessandro Silva for helpful
discussions.  This work was supported by the Lady Davis fund, by the
Israel Science Foundation grant No. 193/02-1 and Centers of
Excellence, and by the German Federal Ministry of Education and
Research (BMBF), within the framework of the German-Israeli Project
Cooperation (DIP).

\end{multicols}


\begin{thebibliography}{10}
\bibitem{detectingnote}
The present derivation of the applicability of Richardson's exact solution to
  the universal Hamiltonian is easily generalized to any set of singly occupied
  levels.
\end{thebibliography}
\end{document}